\documentclass[conference]{IEEEtran}
\usepackage{color}
\usepackage{cite}
\usepackage[normalem]{ulem}
\usepackage{graphicx}
\usepackage{psfrag}
\usepackage{subfigure}
\usepackage{url}
\usepackage{stfloats}
\usepackage{amsmath}
\usepackage{amsthm,mathrsfs,amsfonts,dsfont} 
\usepackage{array}
\usepackage{epsfig}
\usepackage{setspace}
\usepackage{amssymb}
\usepackage{diagbox}
\usepackage{slashbox}

\usepackage{algorithm}
\usepackage{algorithmic}

\usepackage{booktabs}

\usepackage{bm}

\usepackage{comment}

\definecolor{red}{rgb}{0.9,0.1,0.1}
\definecolor{purple}{rgb}{0.625,0.125,0.9375}
\definecolor{brown}{rgb}{0.5,0.3,0.1}
\definecolor{orange}{rgb}{1,0.4,0}
\definecolor{yellow}{rgb}{0.9,0.8,0.2}
\definecolor{green}{rgb}{0.1,0.75,0.1}
\definecolor{cyan}{rgb}{0.1,0.8,0.8}
\definecolor{lblue}{rgb}{0.1,0.4,0.8}
\definecolor{blue}{rgb}{0.1,0.1,0.9}
\definecolor{magenta}{rgb}{0.9,0,0.8}
\definecolor{mgreen}{rgb}{0.2,0.5,0.2}
\definecolor{gray}{rgb}{0.5,0.5,0.5}
\definecolor{reliefcolor}{gray}{0.8}
\definecolor{emphcolor}{rgb}{1,.5,1}
\definecolor{emphcolorr}{rgb}{1,.9,1}
\definecolor{emphcolorb}{rgb}{.9,.9,1}
\definecolor{emphcolorg}{rgb}{.9,1,.9}

\IEEEoverridecommandlockouts\IEEEpubid{\makebox[\columnwidth]{ 978-1-6654-3540-6/22/\$31.00 ~\copyright~2022 IEEE \hfill} \hspace{\columnsep}\makebox[\columnwidth]{ }}

\begin{document}
\title{EMF-Aware MU-MIMO Beamforming in RIS-Aided Cellular Networks}
%\author{\IEEEauthorblockN{Yi Yu$^{(1,2)}$, Lina Mroueh$^{(1)}$, Shuo Li$^{(1)}$ and Michel Terr\'e$^{(2)}$}
\author{\IEEEauthorblockN{Yi Yu, Rita Ibrahim and Dinh-Thuy Phan-Huy}
\IEEEauthorblockA{
 Orange Innovation \\
92320, Châtillon, France \\
%(3) Sequans Communications,  92700 Colombes \\
Email: yu.yi@orange.com, rita.ibrahim@orange.com, dinhthuy.phanhuy@orange.com
}}
\maketitle

\begin{abstract}
Reconfigurable Intelligent Surfaces (RISs) are one of the key emerging 6th Generation (6G) technologies that are expected to improve the link budgets between transmitters and receivers by adding artificial propagation paths. In such re-configured propagation environment, Downlink (DL) Multi-User Multi-Input Multi-Output (MU-MIMO) brings capacity improvement to cellular networks. It benefits from the spatial dimension offered by MIMO systems to enable simultaneous transmission of independent data streams to multiple users on the same radio resources by applying appropriate Beamforming (BF) schemes. However, in some cases, serving the same subset of users for a long period of time may cause some undesired regions where the average Electromagnetic Field Exposure (EMFE) exceeds the regulatory limits. To address this challenge, we propose in this paper a novel Electromagnetic Field (EMF) aware MU-MIMO BF scheme that aims to optimize the overall capacity under EMF constraints in RIS-aided cellular networks.

\end{abstract}


\section{Introduction} 
\label{sec:motivation}

\IEEEPARstart{6}{G} is currently in an early research phase and has substantial potential \cite{OrangeSite}. A number of novel technologies are being investigated as candidates for 6G. The Reconfigurable Intelligent Surface (RIS) is one of these enablers \cite{RIS}. A RIS is a large array of low-cost passive elements, that applies a phase-shift to impinging waves in order to reflect them into desired directions. Thus, these RISs reconfigure the propagation environment by adding artificial propagation paths and in this way improve the Signal-to-Noise Ratio (SNR) at the receiving end.

In upcoming cellular network generation, mobile broadband services will keep-on exploiting Downlink (DL) multi-user multiple-input and multiple-output techniques \cite{pinchera2021optimizing} and  massive MIMO (M-MIMO) antennas to meet the increasing data rate demands. MU-MIMO technique enables the multiplexing of several data streams by applying appropriate BF weights that steer the signal to target devices and suppress partially or completely the impact of interfering streams.  Transmitting independent data streams over parallel spatial channels can increase system capacity and avoid wastage of system resources.

However, in some propagation conditions, the radiation pattern resulting from the DL MU-MIMO BF may create some regrettable regions with strong EMFE. The human exposure to EMF is regulated not to exceed some reference values averaged over a specific time-period \cite{baracca2018statistical}. These regulation limits are usually met thanks to several averaging factors such as resource
utilization, Time-Division Duplex (TDD), scheduling time, and spatial distribution of served users. However, respecting EMF constraints becomes  challenging when the same subset of users is served for a long period of time (e.g. fix wireless access use case). This issue seems to be more exigent in some countries \cite{gsma} that have adopted more stringent EMF limits than those recommended by the International Commission on Non-Ionizing Radiation Protection (ICNIRP) \cite{icnirp}.
%This issue seems even more exigent in some countries \cite{gsma} where stricter EMF limits are adopted compared to the International Commission on Non-Ionizing Radiation Protection (ICNIRP)   recommendation\cite{icnirp}.

Therefore, developing spectral efficient EMF-aware   technologies is necessary to ensure their deployment in future 6G networks. In \cite{awarkeh2021electro,DT1RIS,DT2RIS,xu2019analysis}, the authors have proposed different EMF-aware BF schemes in RIS-aided SU-MIMO networks. In this paper, we focus on RIS-aided MU-MIMO scenario and aim to design efficient BF scheme to optimize the DL network capacity while satisfying EMF constraints. We assume that some RISs
are randomly distributed as reflective surfaces to work on transmitting the incident signal to specified User Equipments (UE). First, we examine a "reference" MU-MIMO BF scheme that maximizes the DL capacity with full power transmission and without any EMF constraint. It corresponds to a Zero-Forcing (ZF) precoding with a water-filling power allocation strategy \cite{caire2003achievable}-\cite{spencer2004zero}. Then, we propose a "reduced" EMF-aware BF which consists of decreasing the overall transmit power until the EMF limits are fulfilled. Moreover, we elaborate a novel "enhanced" EMF-aware BF with a per-layer power control mechanism that is designed to meet EMF constraints and achieve higher capacity performance.

The rest of the paper is organized as following. A MU-MIMO RIS-aided network model is defined in section \ref{sec:model}. Section \ref{sec:reference} introduces the reference BF scheme. Then, we review in Section \ref{sec:beamforming} the two proposed EMF-aware BF schemes. Next, in section \ref{sec:numerical}, the performance evaluation of these BF schemes with respect to DL channel capacity and power efficiency is presented. Finally, section \ref{sec:conclusion} concludes the paper.

\section{System Model}
\label{sec:model} 

\begin{figure}[htbp]%[htbp]
\centering 
\includegraphics[width = 0.56\linewidth]{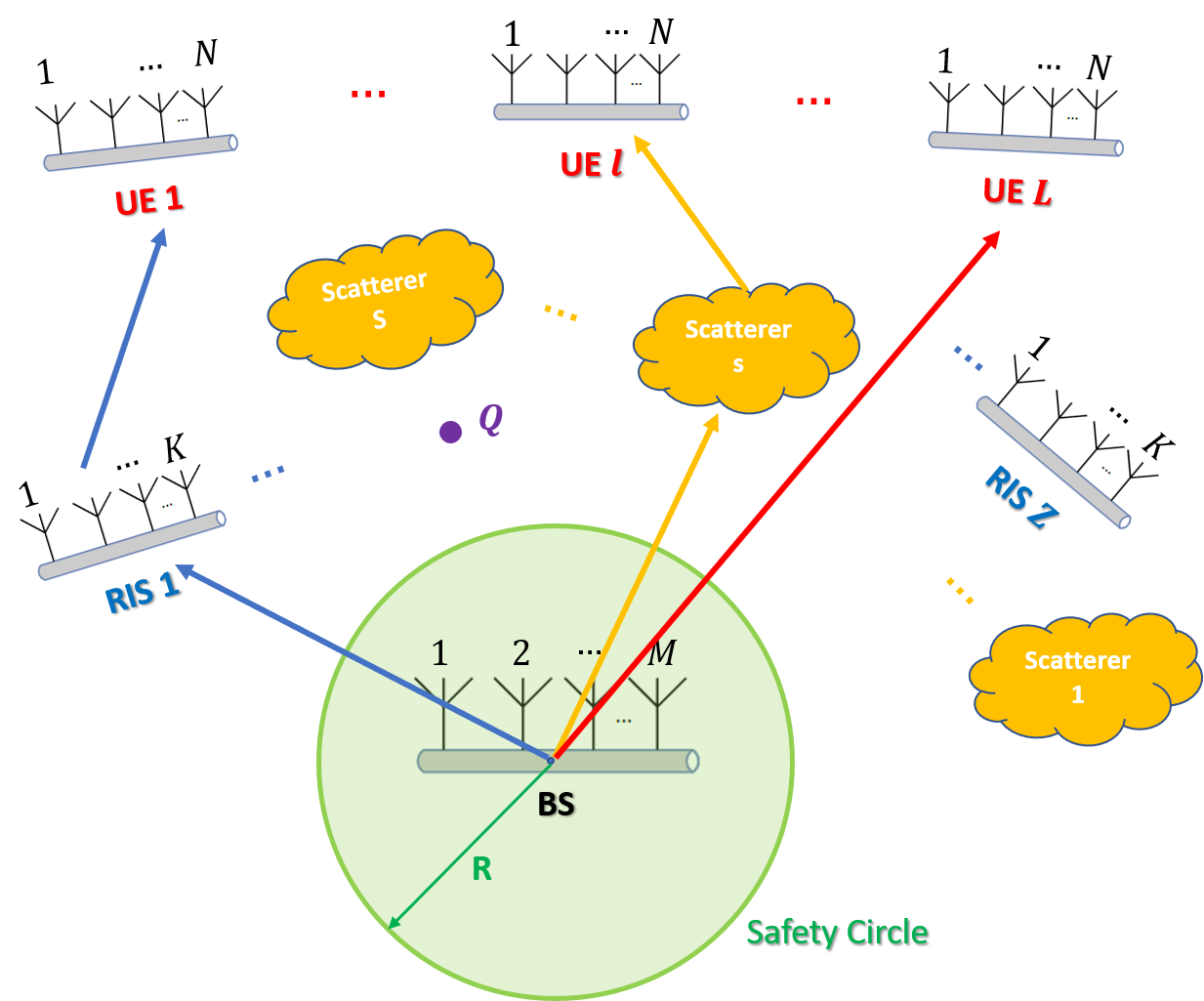}
\caption{A MU-MIMO RIS-aided Network Model}
\label{fig:network}
\end{figure}

% \subsection{MU-MIMO Network Assumptions}
% \label{sec:network}
In this section, we consider the DL communication of a RIS-aided MU-MIMO network. As depicted in Fig. \ref{fig:network}, there are a BS and $L$ different UEs in the cellular network, they are all equipped with multiple antennas.
Assume that the BS has a linear antenna array of $M$ antenna elements and each UE has %a linear array composed of 
$N$ antenna elements. Thus, the total number of received antennas is $N_t=LN$. Those antenna elements are spaced by $0.5\lambda$ to the adjacent ones either on the BS or a single UE side, where $\lambda$ indicates the wavelength of the carrier frequency.
We assume that the BS must serve the UEs for a long period with the same BF scheme and power allocation, during the entire period. Therefore, the EMFE constrains must also be met instantaneously, to be met in average.

%$S \geqslant 1$
There are $S$ scatterers and $Z$ RISs randomly located in the given space, respectively. Each RIS has a linear array of $K$ elements with a spacing of $0.5\lambda$. For simplicity, we consider here the far-field calculation method, i.e., both scatterers and RISs are far from the BS and the UEs. In the far-field, the electromagnetic waves propagate at the speed of light and electric and magnetic fields are mutually perpendicular \cite{baracca2018statistical}. 

We consider an Orthogonal Frequency Division Multiplexing (OFDM) waveform and random Rayleigh fading. With spatial multiplexing, multiple streams are sent from the BS to distinct active UEs simultaneously, which are separated by using precoding schemes. In our case, the ZF linear precoding scheme adapted to multiple receiving antennas is applied. 
%The precoding matrix also refers to the beamformer matrix, denoted by $\mathbf{P}_{zf} = \mathbf{B} \in \mathbb{C}^{M \times \nu}$, where $\nu$ is the number of independent spatial layers.
The network adopts TDD mode and thus the channel reciprocity is feasible. 

In this work, we only focus on the BS BF, where the BS and RIS BF weights will be jointly optimized in our next work phase. First, the BF at the RIS side is selected based on the following procedure: (i) each UE sends some pilots which allow the RIS to estimate the UE-to-RIS channels, (ii) then based on this channel estimation, each RIS computes the BF reflection weight $\mathbf{w}^{z} \in \mathbb{C}^{K \times 1}$, reconfigures the weights and freezes. The phase shift weight is multiplied by a reflection amplitude $r^{ris}$, where $0 \leq r^{\text {ris }} \leq 1$ is a constant value depending on the hardware structure of the RIS. Here we set $r^{r i s}=1 / K$. Once the RIS is configured, the UE sends pilots again, the BS can estimate the DL channel taking into account the RIS configuration and determine the appropriate BS BF to be used for data transmission.
\iffalse
As we know, the BF weight is actually a phase shift. In this work, we do not optimize the BF weights jointly, which can be reached in our next work phase. Furthermore, for the path through a RIS, the incident signal is also multiplied by a reflection amplitude $r^{ris}$, where $0 \leq r^{\text {ris }} \leq 1$ is a constant value depending on the hardware structure of the RIS. Here we set $r^{r i s}=1 / K$.
\fi

In order to satisfy the EMF exposure compliance, a safety circle of radius $R$ centered at the BS is defined, and outside the circle, the received power at any location within the observation range should not exceed a given threshold $\text{EMF}_{\text {th }} \in \mathbb{R}^{+}$. The security circle, also known as the exclusion zone, is guaranteed to be closed to the public.

There are two different kinds of propagation paths, the first one, denoted as $m \rightarrow s \rightarrow U_{n}^{l}$, is from the $m^{th}$ ($1 \leqslant m \leqslant M$) antenna element of the BS to the $n^{th}$ ($1 \leqslant n \leqslant N$) antenna element of the $l^{th}$ ($1 \leqslant l \leqslant L$) UE $U_{n}^{l}$ through scatterer $s$ ($1 \leqslant s \leqslant S$). Another type of propagation path, denoted as $m \rightarrow R^z_k \rightarrow U_{n}^{l}$, is from the $m^{th}$ BS antenna element to the $n^{th}$ antenna element of the $l^{th}$ UE, through $z^{th}$ ($1 \leqslant z\leqslant Z$) RIS's element $R^{z}_k$ ($1 \leqslant k \leqslant K$).

 \begin{figure}[htbp]%[htbp]
 \includegraphics[width = \linewidth]{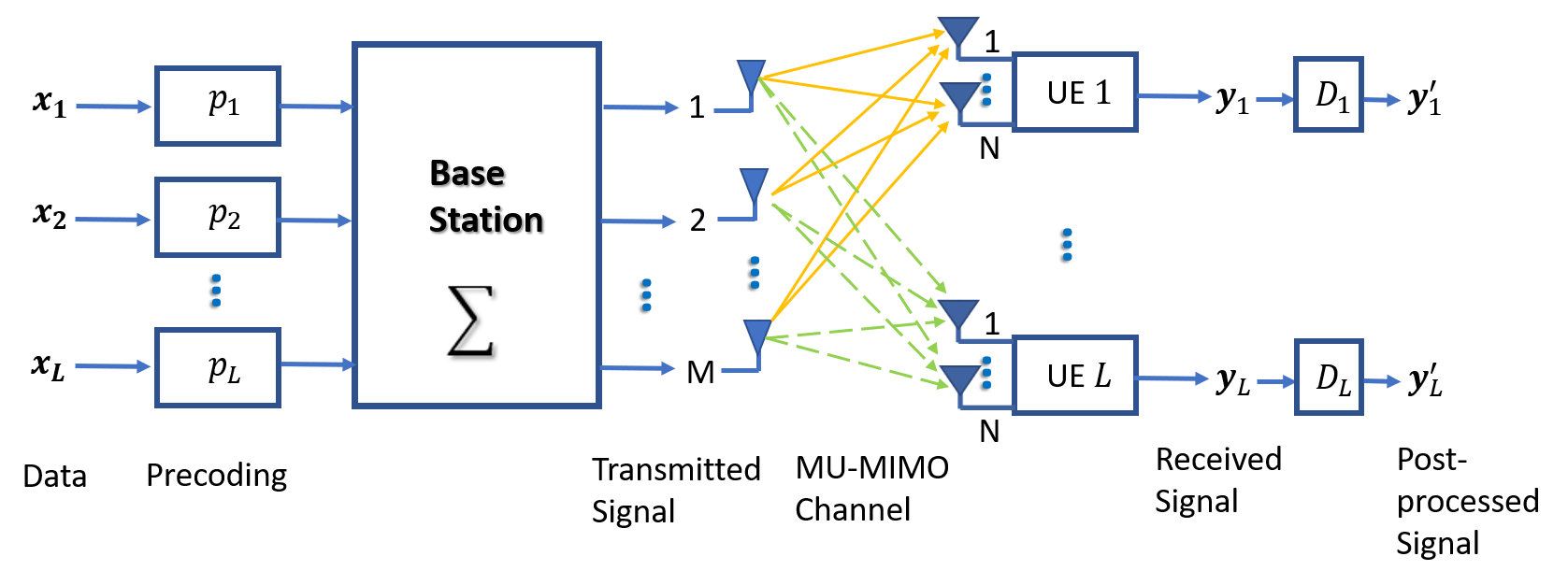}
 \caption{DL MU-MIMO Scheme}
 \label{fig:MIMO}
 \end{figure}

The propagation channel $\boldsymbol{H}_{l} \in \mathbb{C}^{N \times M}$ between the BS and a given UE $l$ through the scatterers and the RISs is given by:

\begin{equation}
\boldsymbol{H}_{l}[n,m]=\sum_{s=1}^{S} G_{m, s, U_{n}^{l}} +\sum_{z=1}^{Z} \sum_{k=1}^{K} G_{m, R_{k}^{z}, U_{n}^{l}},
\end{equation}
where $G_{m, s, U_{n}^{l}}$ and $G_{m, R_{k}^{z}, U_{n}^{l}}$ are the channel gains for paths $m \rightarrow s \rightarrow U_{n}^{l}$ and  $m \rightarrow R^z_k \rightarrow U_{n}^{l}$, respectively. For the calculation of $G_{m, s, U_{n}^{l}}$ and $G_{m, R_{k}^{z}, U_{n}^{l}}$, please refer to Appendix \ref{app:derivation}. 
Hence, the combined channel matrix $\boldsymbol{H}$ is written as:

$$\boldsymbol{H}=\left[\begin{array}{l}\boldsymbol{H}_{1} \\ \boldsymbol{H}_{2} \\ \cdots \\ \boldsymbol{H}_{L}\end{array}\right] \in \mathbb{C}^{N_t \times M}.$$

Assume that the propagation between the BS and a random nearby position $Q \in \mathbb{R}^{3 \times 1}$ is free space propagation,  $\boldsymbol{H}_{m}^{Q} $ is the $m^{th}$ coefficient of the channel model $\boldsymbol{H}^{Q} \in \mathbb{C}^{1 \times M}$.
Here $\boldsymbol{H}_{m}^{Q}$ is calculated by:

\begin{equation}
\boldsymbol{H}_{m}^{Q}=\frac{\lambda e^{j \frac{2 \pi}{\lambda}\left\|\overrightarrow{A^{BS}_{m} Q}\right\|}}{4 \pi\left\|\overrightarrow{A^{BS}_{m} Q}\right\|},
\end{equation}
where $A^{BS}_m\in \mathbb{R}^{3 \times 1}$ is the position of the $m^{th}$ antenna element of the BS.

As shown in Fig \ref{fig:MIMO}, 
%let the MIMO channel of user $l$ is denoted by $\boldsymbol{H}_{l} \in \mathbb{C}^{N \times M}$. 
the data vector $\boldsymbol{x}=\left[\boldsymbol{x}_{1}^T, \boldsymbol{x}_{2}^T, \cdots, \boldsymbol{x}_{L}^T\right]^T \in \mathbb{C}^{\nu \times 1}$  is waiting for transmission,  whereas $\boldsymbol{x}_{l} \in \mathbb{C}^{\nu_l \times 1}$ and $\nu$ is the total number of spatial layers $\nu=\sum_{l=1}^{L} \nu_{l}$. The component of the data vectors are normalized, i.e. $\mathbb{E}\left[{\|\boldsymbol{x}\|}^2\right]=1$. The data $\boldsymbol{x}$ should be pre-processed via the ZF precoder with the BF matrix denoted as $\boldsymbol{B} = [\mathbf{B}_{1}\cdots\mathbf{B}_{L}]\in \mathbb{C}^{M \times \nu}$. %with $\mathbf{B}_{l}\in \mathbb{C}^{M \times \nu_l}$ the precoding matrix of the $l^{th}$ user device. 
Then, the final transmitted signal $\boldsymbol{s} \in \mathbb{C}^{M \times 1}$ is given by:

\begin{equation}
\boldsymbol{s}= \boldsymbol{B}\boldsymbol{x}.
\end{equation} 

The received signal vector is,

\begin{equation}
\boldsymbol{y}= \boldsymbol{H}\boldsymbol{B}\boldsymbol{x}+\boldsymbol{n},
\end{equation}
where $\boldsymbol{n}$ indicates the random receiving noise.

\section{Reference BF Scheme}
\label{sec:reference}

In this section, without considering the EMF constraints, we adopt a ZF precoding based BF scheme with water-filling power control to help the BS to transmit the signals under the total transmit power constraint $P_{max}$. More specifically, since the component of the data vectors are normalized, i.e. $\mathbb{E}\left[\boldsymbol{x}\boldsymbol{x}^{H}\right]=1$, the power constraint is expressed as follows:
\[
\mathbb{E}\left[ {\|\boldsymbol{Bx}\|}^2\right]=tr\left[ \boldsymbol{B}\boldsymbol{B}^{H}\right] \leqslant P_{max},
\]

%We assume a water-filling power allocation per layer, thus $P_t = {P_{max}}/\nu$ for all $t \in \lbrace 1,...,\nu \rbrace$.

The SVD of each full-rank channel matrix $\boldsymbol{H}_{l} \in \mathbb{C}^{N \times M}$ corresponding to UE $l$ is given by: 

\begin{equation}
\boldsymbol{H}_l =\boldsymbol{U}_l\boldsymbol{\Lambda}_l  \boldsymbol{V}_{l}^{H},  
\label{eq:SVD}
\end{equation}
where $\boldsymbol{U}_l\in \mathbb{C}^{N \times N}$ and $\boldsymbol{V}_l\in \mathbb{C}^{M \times N}$ are respectively the unitary orthogonal matrices representing the subset of the left-singular and  right-singular vectors.
$\boldsymbol{V_l}^{H}$ represents the conjugate transpose of $\boldsymbol{V_l}$.
$\boldsymbol{\Lambda}_l=diag\left\{\sqrt{\lambda_{l,1}}, \cdots, \sqrt{\lambda_{l,N}} \right\} $ is a ${N \times N}$ diagonal matrix containing the singular vectors of the channel matrix $\boldsymbol{H}_l$.

Taking into account the receiving diversity at the UE level, the ZF BF matrix $\boldsymbol{B}$ is determined as the pseudo-inverse of the concatenated matrix $\boldsymbol{V}$ defined as follows:

$$\boldsymbol{V} = \left[\mathbf{V}_{1},   \mathbf{V}_{2}, \ldots, \boldsymbol{V}_{L}\right]^H \in \mathbb{C}^{N_t \times M}.$$

%$$\boldsymbol{V} = \left[\mathbf{V}_{1},   \mathbf{V}_{2}, \ldots, \boldsymbol{V}_{L}\right] \in \mathbb{C}^{M \times ML}.$$

The pseudo-inverse $\boldsymbol{V}^{+}\in \mathbb{C}^{M \times N_t}$ of the matrix $\boldsymbol{V}$ is determined by:

\begin{equation}
\boldsymbol{V}^{+} = \boldsymbol{V}^H \left(\boldsymbol{V} \boldsymbol{V}^{H}\right)^{-1}.
\label{eq:pseudo}
\end{equation}

% {\color{red} 
% Proposition of modification:
% $$\boldsymbol{V} = \left[\mathbf{V}_{1},   \mathbf{V}_{2}, \ldots, \boldsymbol{V}_{L}\right]^H \in \mathbb{C}^{ML \times M}.$$

% The pseudo-inverse of the matrix $(\boldsymbol{V})\in \mathbb{C}^{ML \times M}$ is determined by:
% \begin{equation}
% (\boldsymbol{V})^{+} = \boldsymbol{V}^H \left(\boldsymbol{V} \boldsymbol{V}^H\right)^{-1} \in \mathbb{C}^{M \times ML}
% \label{eq:pseudo}
% \end{equation}
% }

In our scenario, we pick only some layers that we are interested in, e.g., $\nu_l \leq rank(\boldsymbol{H_l})$ layers per receiver. Then the total number of layers is equal to $\nu = \sum_{l =1}^{L}\nu_l$ with $\nu \leq N_t$.  With only $\nu$ layers being selected, the matrix $\boldsymbol{V}^{+}$ is trimming to  $\widetilde{\boldsymbol{V}}^{+} \in \mathbb{C}^{M \times \nu}$.
%Then, the beamforming matrix $\boldsymbol{B} \in \mathbb{C}^{M \times \nu}$ is inferred by trimming $\widetilde{\boldsymbol{B}}$.
A total of $\nu$ vertical columns corresponding to the different selection layers are retrieved from $\boldsymbol{V}^+$ and reconstituted into this $\widetilde{\boldsymbol{V}}^+$ matrix.

The BF matrix $\boldsymbol{B}\in \mathbb{C}^{M \times \nu}$ is then deduced as:

\begin{equation}
\begin{aligned}
\boldsymbol{B} =\widetilde{\boldsymbol{V}}^{+}  \boldsymbol{\Sigma}, 
\end{aligned}
\end{equation}
with $\boldsymbol{\Sigma} \in \mathbb{C}^{\nu \times \nu}$ being a diagonal power allocation matrix. 
%In our scenario, for the massive MIMO DL communication, we pick only some layers that we are interested in, e.g., $\nu_l \leq rank(\boldsymbol{H_l})$ layers per receiver are selected. Then the total number of layers is equal to $\nu = \sum_{l =1}^{L}\nu_l$ with $\nu \leq N^t$.  
The transmit power coefficient of a selected layer is set to $P_i$,  where $i\in [1,\cdots,\nu]$. Therefore, $\boldsymbol{\Sigma}$ is denoted as,

\begin{equation}
 \begin{array}{l}
 \boldsymbol{\Sigma}=
 diag\{\underbrace{\sqrt{P_{1}}, \cdots, \sqrt{P_{\nu_1}}}_{\nu_{1}}, \cdots,
 \underbrace{\sqrt{P_{\nu-\nu_L}}, \cdots, \sqrt{P_{\nu}}}_{\nu_{L}} \} .
 \end{array}
 \end{equation}

% \begin{equation}
%  \begin{array}{l}
%  \boldsymbol{\Sigma}=
%  diag\{\underbrace{\sqrt{P_{1}}, \cdots, \sqrt{P_{\nu_1}}}_{\nu_{1}},\quad \underbrace{0, \cdots, 0}_{N-\nu_{1}}, \quad \cdots,\\ \qquad \qquad \qquad
%  \underbrace{\sqrt{P_{\nu-\nu_L}}, \cdots, \sqrt{P_{\nu}}}_{\nu_{L}} , \quad \underbrace{0, \cdots, 0}_{N-\nu_{L}}\} .
%  \end{array}
%  \end{equation}

%Then, the beamforming matrix $\boldsymbol{B} \in \mathbb{C}^{M \times \nu}$ is inferred by trimming $\widetilde{\boldsymbol{B}}$. That is, a total of $\nu$ vertical columns representing the different selection layers are retrieved from $\widetilde{\boldsymbol{B}}$ and reconstituted into this $\boldsymbol{B}$ matrix. Obviously, the values in the columns corresponding to the unselected layers in matrix $\widetilde{\boldsymbol{B}}$ are all $0$.

%$$\boldsymbol{H}=\left[\begin{array}{l}\boldsymbol{H}_{1} \\ \boldsymbol{H}_{2} \\ \cdots \\ \boldsymbol{H}_{L}\end{array}\right] \in \mathbb{C}^{N^t \times M}.$$

%  \begin{equation}
%  \boldsymbol{\Sigma_{\nu}}=diag\{\underbrace{\frac{1}{\alpha}, \cdots, \frac{1}{\alpha}}_{\nu_{1}}, \underbrace{0, \cdots, 0}_{N-\nu_{1}}, \cdots, \underbrace{\frac{1}{\alpha}, \cdots, \frac{1}{\alpha}}_{\mathcal{\nu}_{L}}, \underbrace{0, \cdots, 0}_{N-\nu_{L}}\}.
%  \end{equation}

%  \begin{equation}
%  \begin{aligned}
%  &\boldsymbol{\Sigma}= \\& diag\{ \underbrace{\sqrt{P_1}, \cdots, \sqrt{P_{\nu_1}}}_{\nu_{1}},\cdots, \underbrace{\sqrt{P_{\nu-\nu_L}}, \cdots,  \sqrt{P_{\nu}}}_{\mathcal{\nu}_{L}}\}.
%  \end{aligned}
% \end{equation}

As mentioned previously, the total transmit power is bounded by $tr\left[ \boldsymbol{B}\boldsymbol{B}^{H}\right]\leqslant P_{max}$. So we have, 
% \begin{equation*}
% \begin{aligned}
% tr\left[ \boldsymbol{B}\boldsymbol{B}^{H}\right] = tr\left[ \widetilde{\boldsymbol{B}}\widetilde{\boldsymbol{B}}^{H}\right]\leq P_{max} 
% \end{aligned}
% \end{equation*}

$$
 tr\left[ \boldsymbol{B}\boldsymbol{B}^{H}\right] = tr\left[ \widetilde{\boldsymbol{V}}^{+}  \boldsymbol{\Sigma}  \left(\widetilde{\boldsymbol{V}}^{+}  \boldsymbol{\Sigma}\right)^H  \right] \leqslant P_{max};
$$

$$
\Rightarrow tr\left[\boldsymbol{\Sigma}^2    \left(\widetilde{\boldsymbol{V}}\widetilde{\boldsymbol{V}}^{H}\right)^{-1}  \right] \leqslant P_{max}.
$$

Thanks to ZF precoding, the interference between different users is eliminated and the DL capacity of the MU-MIMO system is approximated by:
\iffalse
\begin{equation}
C=\sum_{i=1}^{\nu} \log \left(1+\frac{\left|\boldsymbol{H} \boldsymbol{B}\right|_i^{2}}{N_{0}}\right) \text { bit/symbols }
\end{equation}
\fi
\begin{equation}
C=B_0 \, \sum_{i=1}^{\nu} \log \left(1+\frac{\lambda_i  P_i}{N_{0}}\right),
\text { Mbits\,/\,s},
%\text { bit\,/\,s\,/\,Hz},
\end{equation}
where  $B_0$ represents the bandwidth and $N_0$  is  the  power  density  of  the  noise.  

To achieve the maximum data rate, we are going to find the transmit power allocation that satisfies this optimization expression:

\begin{equation}
%\mathrm{C}_{N}
\mathrm{C}^{*}:=\max _{P_{1}, \ldots, P_{\nu}} B_0 \, \sum_{i=1}^{\nu} \log \left(1+ \frac{\lambda_i P_i}{N_{0}}\right),
\label{eq:dataRate}
\end{equation}

% \[ 
% \text{s.t.}\,\,\,\, tr\left[\boldsymbol{\Sigma}^2   \left(\boldsymbol{H}\boldsymbol{H}^H\right)^{-1}  \right] = P_{max}
% \]

\[ 
\text{s.t.}\,\,\,\, tr\left[\boldsymbol{\Sigma}^2    \left(\widetilde{\boldsymbol{V}}\widetilde{\boldsymbol{V}}^{H}\right)^{-1}  \right] = P_{max};
\]

$$
P_i \geqslant 0, \, \, i = 1, \cdots, \nu.
$$

As eq.\ref{eq:dataRate} is a convex problem, the optimal solution satisfying the Karush-Kuhn-Tucker (KKT) conditions is resolvable. We can solve this optimization problem via a water-filling algorithm. The optimal solution $P_i$ can be find as,

\begin{equation}
 P_i = max\left(\frac{1}{\mu  \cdot \left[  \left(\widetilde{\boldsymbol{V}}\widetilde{\boldsymbol{V}}^{H}\right)^{-1}\right]_{ii}} - \frac{N_0}{\lambda_i}, \quad 0\right).
\label{eq:result1}
\end{equation}
where $\mu$ is a non-negative Lagrange multiplier deduced from the derivation of the Lagrangian expression,

\begin{equation}
\mu=\frac{\nu}{P_{\max }+\sum_{i=1}^{\nu} \frac{N_{0} \cdot\left[\left(\widetilde{V} \widetilde{V}^{H}\right)^{-1}\right]_{i i}}{\lambda_{i}}}.
\end{equation}

The received power $P_{Q}$ at a random position $Q$ which is in proximity to the BS, is computed as:

\begin{equation}
P_Q =|\boldsymbol{H}^{Q}  \boldsymbol{B}|^{2}.
\end{equation}

In this mechanism, as we only consider the transmit power constraint, there may be several transmit beams which exceed the EMF threshold out of the safety circle.  In the sequel, two EMF-aware MU-MIMO BF schemes are proposed.

% \begin{table}[h]
% \caption{SINR threshold $\gamma_{SF}$ with sub-bandwidth $B=125$ kHz} %to target symbol error probability of $10^{-4}$}
% \renewcommand{\arraystretch}{1.4}
% \centering
%  \begin{tabular}{|c|c|c|c|c|c|c|}
% \hline
% %\diagbox{$B$}{$SF$}& 7 & 8 & 9 & 10 & 11 & 12\\
% $SF$ & 7 & 8 & 9 & 10 & 11 & 12\\
% %\backslashbox{$B$}{$\gamma$(dB)}{$SF$} & 7 & 8 & 9 & 10 & 11 & 12\\
%  \hline
%  $\gamma_{SF}$ (dB) & $-7.5$ & $-10$ &  $-12.5$ & $-15$ & $-18$ &  $-21$  \\
%  \hline
% % $\rho$ (km) & $11.43$ & $13.9$ &  $17$ & $20.6$ & $23.5$ &  $26.8$  \\
% %  \hline
%  \end{tabular}
%  \label{tab1}
%   \end{table}

\section{EMF-aware MU-MIMO Beamforming in RIS-aided cellular networks}
\label{sec:beamforming}

In this section, we propose two EMF-aware BF schemes for wireless MU-MIMO DL communications, taking into account the power and EMF constraints. The general problem is described as follows:
\begin{equation}
\mathrm{C}^{*}:=\max _{P_{1}, \ldots, P_{\nu}} B_0 \, \sum_{i=1}^{\nu} \log \left(1+ \frac{\lambda_i P_i}{N_{0}}\right),
\label{eq:dataRate1}
\end{equation}

\begin{equation*}
    \text { s.t. }{tr}\left[\boldsymbol{\Sigma}^{2}\left(\widetilde{\boldsymbol{V}} \widetilde{\boldsymbol{V}}^{H}\right)^{-1}\right] \leqslant P_{max};
\end{equation*}
\begin{equation*}
    P_{Q}= tr\left[\boldsymbol{\Sigma}^2( \boldsymbol{H}^{Q} \widetilde{\boldsymbol{V}}^+ )^H(\boldsymbol{H}^{Q} \widetilde{\boldsymbol{V}}^+)\right] \leqslant \text{EMF}_{th}, \,\, Q \in \Omega;
\end{equation*}
%$$ \forall{a \in [1,\cdots,N_Q]};$$
$$
P_i \geqslant 0, \, \, i = 1, \cdots, \nu.
$$
where $\Omega$ is the set of all sampling positions $Q$ on the safety circle.
%{\color{red} to be replaced by optimization problem such that the one described by equation 11}

\subsection{Reduced EMF-aware BF Scheme}

The reduced EMF-aware BF scheme is carried out by using a reduction in the total transmit power of the reference BF. The corresponding reduction factor $\alpha$ is determined by,

% \begin{equation}
% \alpha=min(\frac{\text{EMF}_{\text {th }}  }{max\limits_{Q\in \Omega}(\mathbf{P}_r,\mathbf{P}_Q)},1),
% \end{equation}
% where $\mathbf{P}_r = [P_{r,1},\cdots,P_{r,L}] \in \mathbb{C}^{L \times 1}$ and $\mathbf{P}_Q$ is the set of received power at all sampling points $Q$ on the safety circle near the BS.

\begin{equation}
\alpha=min(\frac{\text{EMF}_{\text {th }}  }{\max\limits_{Q \in \Omega}(\mathbf{P}_Q)},1),
\end{equation}
where  $\Omega$ is the set of all sampling positions $Q$ on the safety circle near the BS and $\mathbf{P}_Q$ is the received power at this position $Q \in \omega$.

Consequently, for the reduced EMF-aware BF, the transmit power per layer is reduced by this factor $\alpha$, the power allocation is given by $\boldsymbol{\Sigma}_{red} = \sqrt{\alpha} \boldsymbol{\Sigma} $ and the total transmit power is equal to $P_{red}  = tr\left[\boldsymbol{\Sigma}_{red}^2    \left(\widetilde{\boldsymbol{V}}\widetilde{\boldsymbol{V}}^{H}\right)^{-1}  \right] $.

By this way, the reduced EMF-aware BF scheme,  denoted as $\boldsymbol{B}_{red}$, fulfills the EMF exposure constraints at the expense of some network capacity and is given by:
\begin{equation}
    \boldsymbol{B}_{red} = \widetilde{\boldsymbol{V}}^{+}  \boldsymbol{\Sigma}_{red}= \sqrt{min(\frac{\text{EMF}_{\text {th }}  }{\max\limits_{Q \in \Omega}(\mathbf{P}_Q)},1)}\cdot \widetilde{\boldsymbol{V}}^{+}  \boldsymbol{\Sigma}.
\end{equation}

\subsection{Enhanced EMF-aware BF Scheme}
\label{sec:truncated}

In this subsection, we propose an enhanced EMF-aware BF scheme $\boldsymbol{B}_{enh}$. Our objective is to find the contribution of each layer to the received power over the safety circle sampling points and to selectively reduce the power of each layer in an iterative way.

First, we sample $N_Q$ points on the safety circle and calculate the received power $P_Q$ of these points. For the point $Q_{max}$ with the highest received power $P_{Q_{max}}$, we find the layer $i_0$ which has the greatest influence on its received power. A reduction factor $\beta_{i_0} = (\text{EMF}_{\text{th}}/P_{Q_{max}})$  is applied to specifically reduce the corresponding transmitted power. \iffalse The enhanced EMF-aware BF scheme $\boldsymbol{B}_{enh}$ is then updated.\fi After that, the received power is recalculated for all sampled points, repeating the above steps until the received power at all sampled points satisfies the EMF limits. The details of this BF scheme are shown in Algorithm \ref{alg:truncated}.

With enhanced EMF-aware BF, we can tightly control the received power at random points on the safety circle to meet the EMF constraints, and have better performance in terms of network capacity compared to the reduced EMF-aware BF scheme. 
%As we dynamically consider the impact of each layer of transmit power on the received power at each sampling point, the total transmit power reduction is relatively smaller. 
A detailed comparison will be presented in next section.

\begin{algorithm}
	\renewcommand{\algorithmicrequire}{\textbf{Input:}}
	\renewcommand{\algorithmicensure}{\textbf{Output:}}
	\caption{Enhanced EMF-aware BF}
	\label{alg:1}
	\begin{algorithmic}[1]
		%\REQUIRE Positions of $N$ active sensor nodes 
     	%\ENSURE SF and transmit power values for each active sensors
     	\STATE Initialize the enhanced EMF-aware BF scheme equal to the reference one, $\boldsymbol{B}_{enh} = \boldsymbol{B}$, so $P_{enh,i} = P_i$ where $i\in [1,\cdots,\nu]$ indicates layers;
     	\STATE 
     	Sample $N_Q$ points uniformly on the safety circle to find the maximum received power $P_{Q_{max} }=\max \left|\boldsymbol{H}^{Q} \boldsymbol{B}_{\text {enh }}\right|^{2}$, and determine the position of this point $Q_{max}= \underset{Q}{\operatorname{argmax}} P_{Q}$;\\

		\WHILE{$P_{Q_{max}}>\text{EMF}_{\text{th}}$}
		\STATE Calculate the impact of different layers on this $Q_{max}$ point's received power, $P_{Q_{max}}(i) =|\boldsymbol{H}^{Q_a}  \boldsymbol{B_{enh}}|_i^{2}$. 
		Find out the layer $i_0$  which has the maximum influence on $P_{Q_{max}}$, $i_0 = \underset{i}{\operatorname{argmax}} P_{Q_{max}}(i)$;\\

		\STATE Reduce the transmit power on this corresponding layer $i_0$, $P_{enh,i_0} = (\text{EMF}_{\text{th}}/P_{Q_{max}}) P_{enh,i_0}$, so $\boldsymbol{\Sigma}_{enh} = diag\{\sqrt{P_{enh,1}}, \cdots,  \sqrt{P_{enh,\nu} }\} $; \\
		\STATE Apply the new enhanced EMF-aware BF scheme $\boldsymbol{B}_{enh} = \widetilde{\boldsymbol{V}}^{+}  \boldsymbol{\Sigma}_{enh} $;
		\STATE Find the new maximum received power $P_{Q_{max} }=\max \left|\boldsymbol{H}^{Q} \boldsymbol{B}_{\text {enh }}\right|^{2}$, and determine the position $Q_{max}$ of this point.
		\ENDWHILE
	    \RETURN $\boldsymbol{B}_{enh}$
	\end{algorithmic} 
	\label{alg:truncated}
\end{algorithm}

\section{Numerical results}
\label{sec:numerical}
In this section, we numerically evaluate the performance of the reduced and the enhanced EMF-aware BF schemes. Assume that the BS is equipped with a linear antenna array at the origin with $M = 64$ elements.  $Z=3$ RIS and $S = 3$ scatterers are randomly distributed in the cellular network. Each RIS has $K = 4$ antenna elements.  There are $L$ number of UEs with random positions in the cell and each UE has $N=4$ antennas. The maximum transmit power of the BS is $P_{max} = 200$ Watt. We set the radius of the safety circle to $R = 50$ m. The EMF-threshold is $\text{EMF}_{th} = -5$ dBm. The carrier frequency is assumed to be $3.5$ GHz and with a channel bandwidth of $100$ MHz.

 \begin{figure}[htbp]
 \centering
   \includegraphics[width=0.44\textwidth,height=0.78\textwidth]{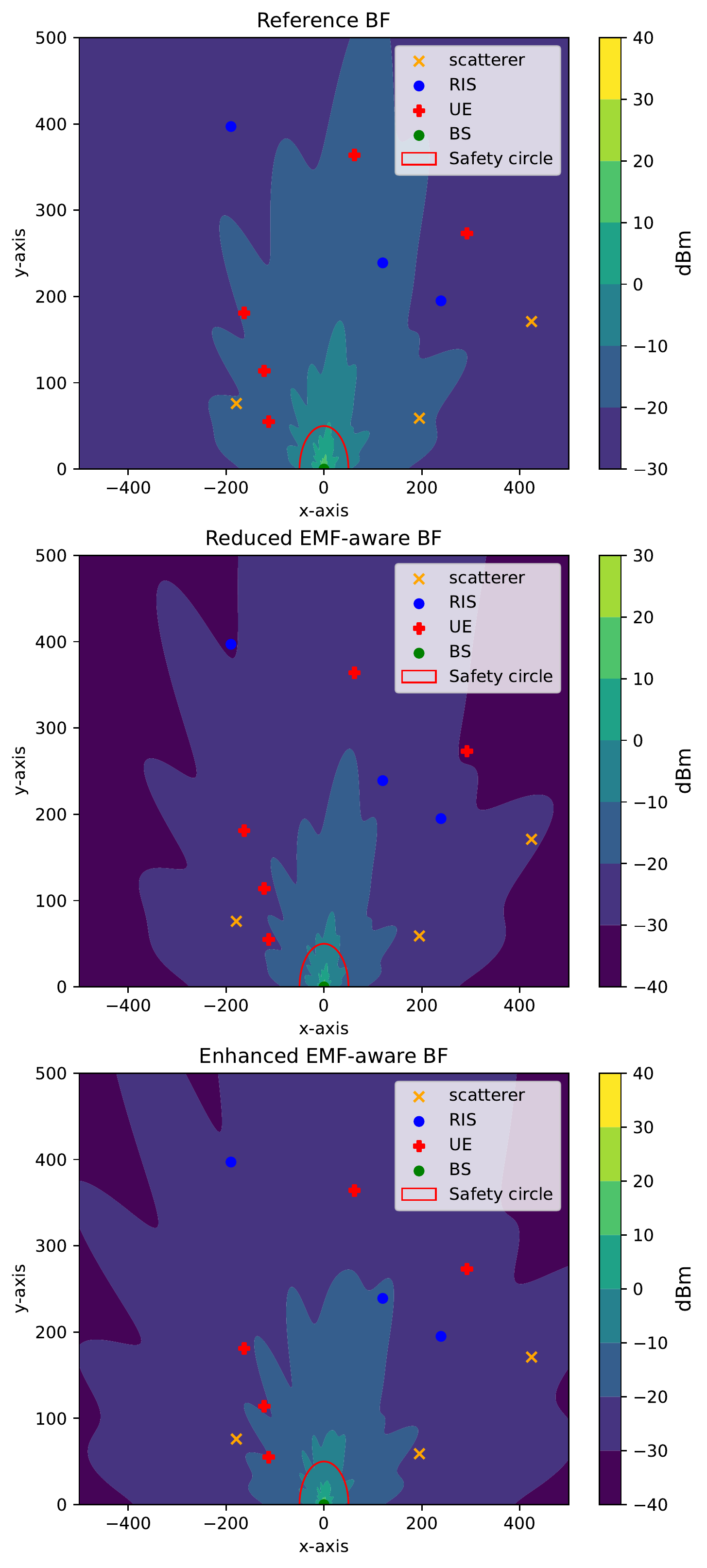} 
           \caption{The received power distribution in a given space with different BF schemes}
               \label{fig:Pr_three}
 \end{figure} 

Figure \ref{fig:Pr_three} shows the received power in a given observation space for $L=5$ UEs. In reference BF case, the received power in the given space ranges from $-30$ dBm to $40$ dBm. By using the reduced and the enhanced EMF-aware BF schemes respectively, the distribution of the received power over the whole given area has been significantly changed.

 \begin{figure}[htbp]
 \centering
   \includegraphics[width=0.37\textwidth,height=0.78\textwidth]{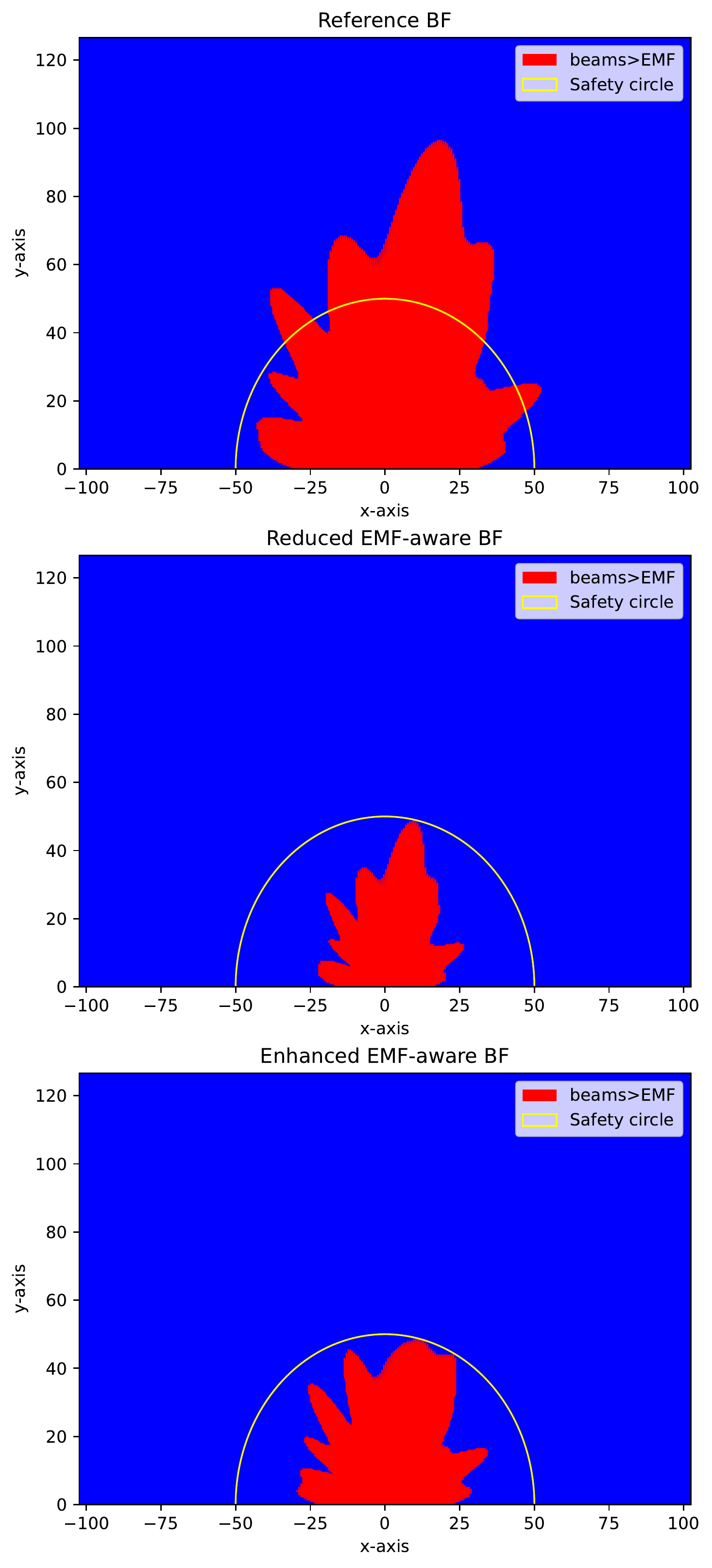} 
           \caption{Beams exceed the EMF constraint in the given space}
               \label{fig:Pr_three_binary}
 \end{figure} 

Figure \ref{fig:Pr_three_binary} is the illustration of beams that exceed the EMF threshold in the same scenario as shown in figure \ref{fig:Pr_three}. 
In the reference case, there are multiple beams that exceed the EMF limits beyond the safety circle. For both EMF-aware BF schemes, the EMF constraints are well adhered in the open space outside the safety circle. The reduced EMF-aware BF achieves this goal by decreasing the overall power by a given factor.  In contrast, in the enhanced algorithm, the transmit power of the different layers is modified in such a way that the exact EMF limits are achieved at the safety circle points that correspond to the exceeding directions.
 
\begin{figure}[htbp]
\centering
\includegraphics[width=0.42\textwidth,height=0.35\textwidth]{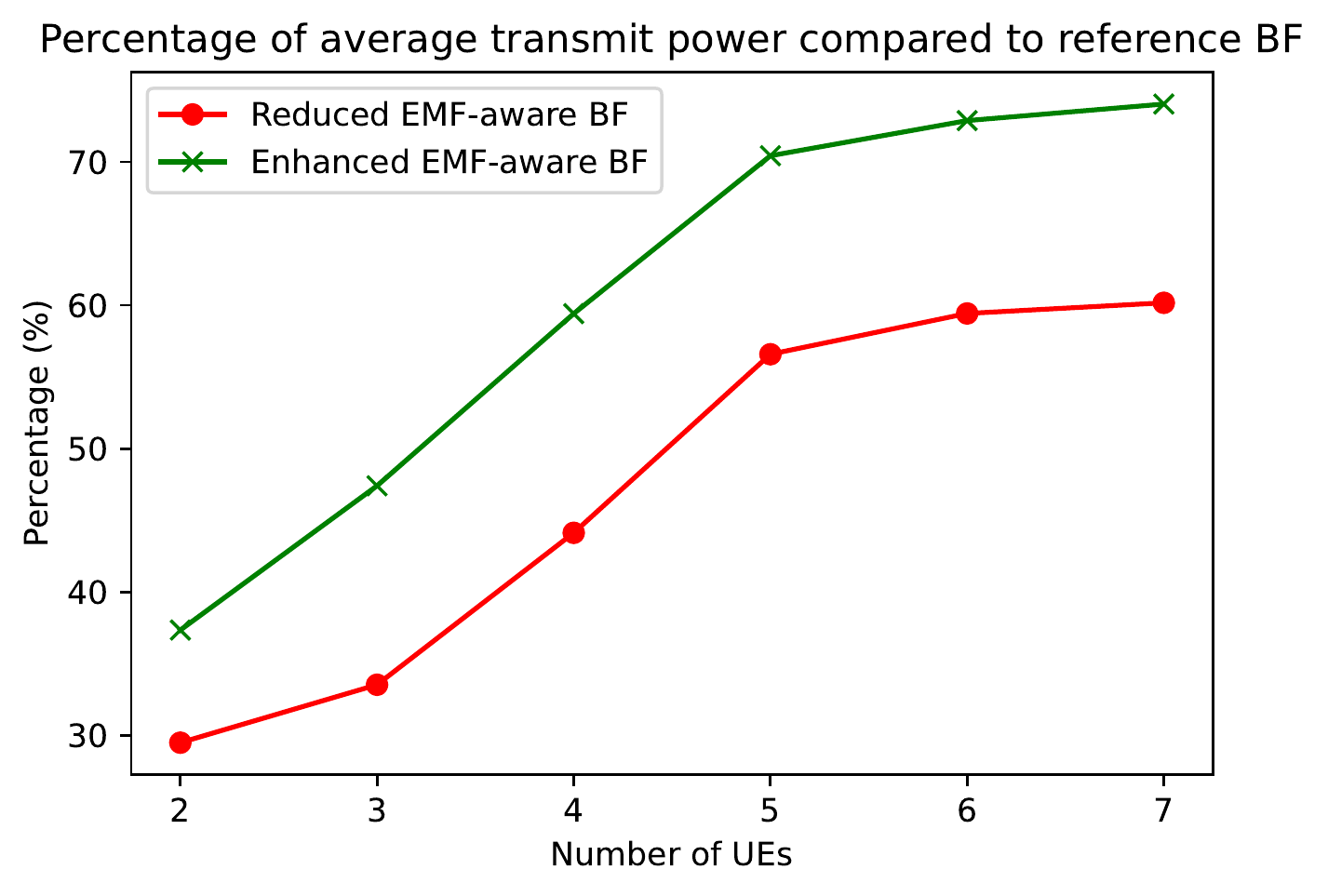} 
\caption{Percentage of average transmit power compared to the reference BF}
\label{fig:average_Pi}
\end{figure}   

\begin{figure}[htbp]%htbp
\centering
\includegraphics[width=0.4\textwidth,height=0.35\textwidth]{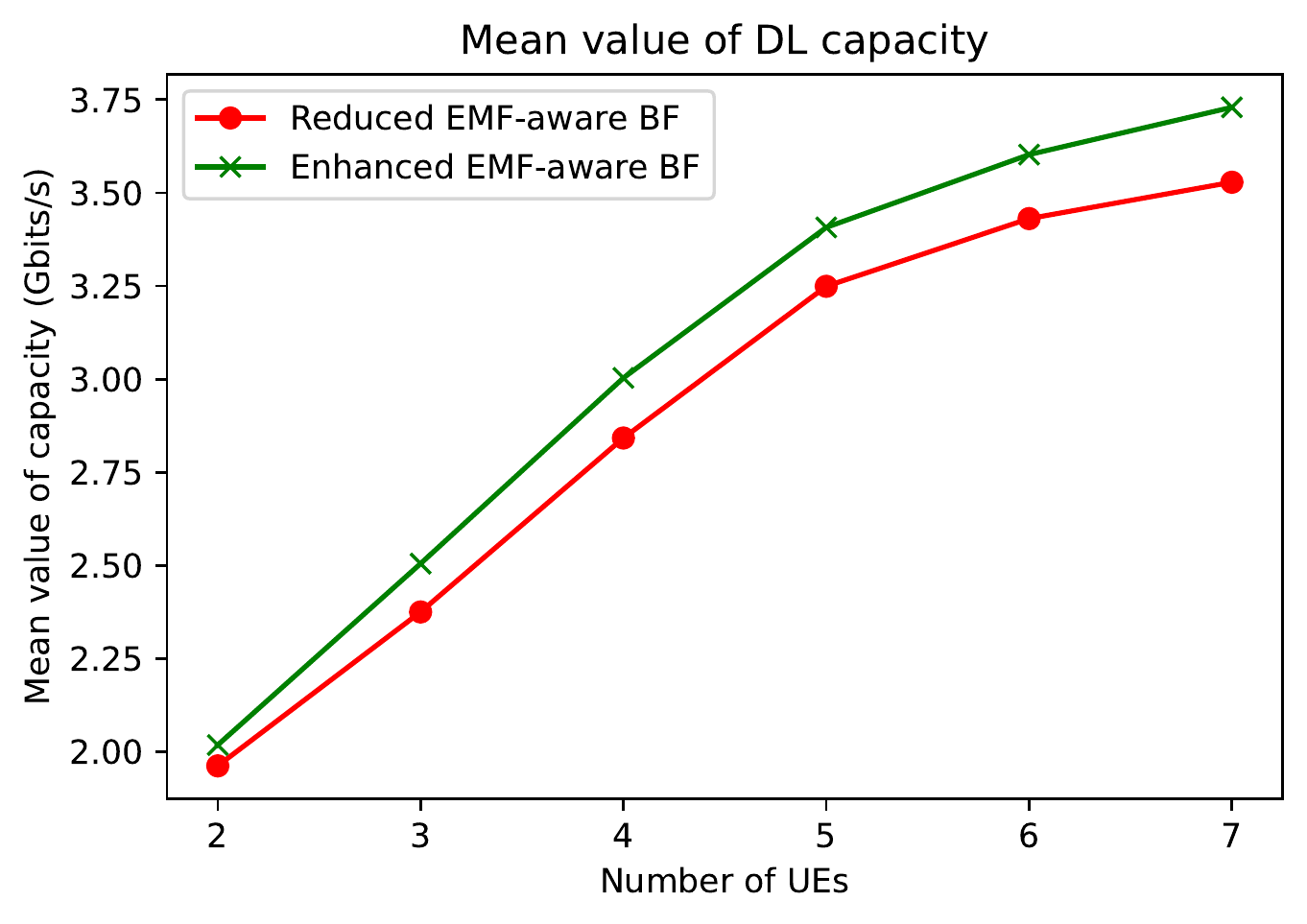} 
\caption{The average DL capacity of the cellular network}
\label{fig:average_c}
\end{figure} 

% \begin{figure}[htbp]
% \centering
% \includegraphics[width=0.42\textwidth,height=0.37\textwidth]{Mean_c_rate1.pdf} 
% \caption{The average DL capacity of the cellular network}
% \label{fig:average_c}
% \end{figure}

Moreover, we evaluate the performance of the proposed BF schemes considering different number of UEs, i.e. from $L=2$ to $7$. We consider $1000$ samples of channels corresponding to each number $L$ in the simulation.
%{\color{red} how many samples of channel are considered for each case?}
Figure \ref{fig:average_Pi} presents the percentage of average transmit power at the BS for the two proposed BF schemes compared to the reference BF. The enhanced EMF-aware BF can still guarantee the EMF limits with about $10\%$ higher transmit power than the reduced one.
%plots the percentage of their transmit power in relation to that of the reference scheme.
\iffalse Figure \ref{fig:average_c} plots the average DL capacity of the cellular network with respect to different BF schemes. The enhanced EMF-aware BF can maintain more than $70\%$ of the capacity, which is $7\%$ higher than the reduced BF. %As can be seen from the figures, the enhanced EMF-aware BF has a higher average transmit power relative to the reduced one and maintains more of the system capacity. %Both BF schemes satisfy the power and EMF constraints.
\fi
Figure \ref{fig:average_c} plots the average DL capacity of the cellular network with respect to different BF schemes. The enhanced EMF-aware BF can maintain more than $70\%$ of the capacity, which is $7\%$ higher than the reduced BF.
\section{Conclusion}
\label{sec:conclusion}

In this paper, we modeled the  DL communcation in a RIS-aided MU-MIMO systems.
Two BF schemes are proposed to address EMF exposure regulation: (i) reduced and (ii) enhanced EMF-aware MU-MIMO BF. We compare the simulation performance of these two schemes. The enhanced EMF-aware scheme achieves a higher system capacity compared to the reduced one which has a lower average transmit power. In the near future, we will jointly optimize the transmit precoding weight and the power allocation scheme in order to achieve higher capacity performance while satisfying EMF regulation.

\section{Acknowledgement}
\label{sec:ack}

This work was conducted within the framework of the European Union's innovation project RISE6G.

%\appendix
%\section{Appendix}
%\label{app:derivation}
\begin{appendices}

\section{}
\label{app:derivation}

As mentioned in section \ref{sec:model}, $G_{m, s, U_{n}^{l}}$ and $ G_{m, R_{k}^{z}, U_{n}^{l}} $ are the channel gains with respect to different propagation paths $m \rightarrow s \rightarrow U_{n}^{l}$ and $m \rightarrow R^z_k \rightarrow U_{n}^{l}$. They are calculated as:

\begin{equation}
G_{m, s, U_{n}^{l}}=\beta(s) \cdot e^{-j \frac{2 \pi}{\lambda}\left(\delta(m, s)+\delta\left(s, U_{n}^{l}\right)\right)} ;
\end{equation}
and \begin{equation}
\begin{aligned}
& G_{m, R_{k}^{z}, U_{n}^{l}}  =\\  & r^{\text {ris }} \cdot \epsilon\left(R_{0}^{z}\right) \cdot e^{-j \frac{2 \pi}{\lambda} \cdot \eta\left(m, R_{k}^{z}\right)} \cdot \mathbf{w}_{k}^{z} \cdot e^{-j \frac{2 \pi}{\lambda} \eta\left(R_{k}^{z}, U_{n}^{l}\right)};
\end{aligned}
\end{equation}
where $\beta\left(s\right)$ and $\epsilon(R^{z}_0)$ are complex random Gaussian variables with unit expectation. $r^{r i s}=1 / K$ is the refection amplitude and  $\mathbf{w}^{z} \in \mathbb{C}^{K \times 1}$ is the RIS BF reflection weight. In addition, \begin{equation}
\delta(m, s)=\frac{\overrightarrow{A_{0}^{BS} A_{s}^{s c a}}}{\left\|\overrightarrow{A_{0}^{BS}  A_{s}^{s c a}}\right\|} \cdot \overrightarrow{A_{0}^{BS}  A_{m}^{B S}};
\end{equation}

%and
\begin{equation}
\delta\left(s, U_{n}^{l}\right)=\frac{\overrightarrow{A_{s}^{s c a}  A_{l, 0}^{UE}}}{\left\|\overrightarrow{A_{s}^{s c a}  A_{l, 0}^{UE}}\right\|} \cdot \overrightarrow{A_{l, 0}^{UE}  A_{l, n}^{U E}};
\end{equation}

\begin{equation}
\eta\left(m, R_{k}^{z}\right)=\frac{\overrightarrow{A_{0}^{B S} A_{z, 0}^{RIS}}}{\left\|\overrightarrow{A_{0}^{BS}  A_{z, 0}^{RIS}}\right\|} \cdot\left(\overrightarrow{A_{0}^{BS}  A_{m}^{BS}}+\overrightarrow{A_{z, 0}^{RIS} A_{z, k}^{RIS}}\right);
\end{equation}

\begin{equation}
\eta\left(R_{k}^{z}, U_{n}^{l}\right)=\frac{\overrightarrow{A_{z, 0}^{RIS} A_{l, 0}^{U E}}}{\left\|\overrightarrow{A_{z, 0}^{RIS}  A_{l, 0}^{U E}}\right\|}\cdot \left(\overrightarrow{A_{z, 0}^{RIS}  A_{z, k}^{RIS}}+\overrightarrow{A_{l, 0}^{U E}  A_{l, n}^{U E}}\right).
\end{equation}
with $A^{BS}_0$ being the center position of the BS linear array, $A^{BS}_m\in \mathbb{R}^{3 \times 1}$ is the position of the $m^{th}$ BS antenna element. Similarly, $A^{UE}_{l,n}\in \mathbb{R}^{3 \times 1}$ is the position of the $n^{th}$ antenna element of the $l^{th}$ UE and $A^{RIS}_{z,k}\in \mathbb{R}^{3 \times 1}$ is the position of the $k^{th}$ element of the $z^{th}$ RIS. The position of scatterer $s$ is denoted as $A^{sca}_{s}\in \mathbb{R}^{3 \times 1}$.
\end{appendices}

\bibliographystyle{IEEEtran}
\bibliography{ref.bib}

\end{document}